\documentclass[12pt]{article}
\usepackage[dvips]{epsfig}
\usepackage{graphpap}

\def\Journal#1#2#3#4#5#6{(#5) ``#6'' {#1} {\bf #2} #3#4}

\newcommand{\bm}[1]{\mbox{\boldmath $#1$}}

\def\CQG{\em Class. Quantum Grav.}
\def\JPA{\em J. Phys. A: Math. Gen.}
\def\PRD{\em Phys. Rev. D }
\def\GRG{\em Gen. Rel. Grav.}

\def\MNRAS{\em Mon. Not. Roy. Astr. Soc.}
\def\JMP{\em J. Math. Phys.}

\def\CMP{\em Commun. Math. Phys.}

\def\d{{\rm \mbox{d}}}

\def\pp{\varphi}

\def\be{\begin{equation}}
\def\ee{\end{equation}}
\def\bea{\begin{eqnarray}}
\def\eea{\end{eqnarray}}
\def\bean{\begin{eqnarray*}}
\def\eean{\end{eqnarray*}}

\newcounter{eqlletra}

\begin{document}

\title{Cylindrically symmetric dust spacetime}

\author{
Jos\'e M. M. Senovilla\thanks{Also at
Laboratori de F\'{\i}sica Matem\`atica,
Societat Catalana de F\'{\i}sica, IEC, Barcelona.} \footnotemark[3]
and
Ra\"ul Vera\footnotemark[1] \footnotemark[4] \\
\footnotemark[3] Fisika Teorikoaren Saila, Euskal Herriko Unibertsitatea,\\
644 P.K., 48080 Bilbao, Spain.\\
\footnotemark[4] School of Mathematical Sciences,
Queen Mary and Westfield College\\ Mile End Road, London E1 4NS, England.\\}

\maketitle

\begin{abstract}
We present an explicit exact solution of Einstein's equations for an 
inhomogeneous dust universe with cylindrical symmetry.
The spacetime is extremely simple but nonetheless 
it has new surprising features. The universe is ``closed'' in the sense
that the dust expands from a big-bang singularity but recollapses to a
big-crunch singularity. In fact, both singularities are connected so that the 
whole spacetime is ``enclosed'' within a single singularity of general 
character. The big-bang is not simultaneous for the dust, and in fact 
the age of the universe as measured by the dust particles depends on
the spatial position, an effect due to the inhomogeneity, and 
their total lifetime has no non-zero lower limit. 
Part of the big-crunch singularity is naked.
The metric depends on a parameter and contains flat 
spacetime as a non-singular particular case. For appropriate values of the 
parameter the spacetime is a small perturbation of Minkowski 
spacetime. This seems to indicate that flat spacetime may be 
unstable against some global {\it non-vacuum} perturbations.
\end{abstract}

Exact solutions of Einstein's equations for an inhomogeneous dust 
with an Abelian $G_{2}$ on $S_{2}$ group of motions seem to be 
difficult to find, and very few of them are known, see \cite{Dust} 
and references therein.
Among them, there are some Petrov type-D known cases 
belonging to the general class found by 
Szekeres \cite{SZE}, some examples appear in \cite{BOTO,WW1}. 
However, it seems that no exact solution for an inhomogeneous dust with 
{\it cylindrical symmetry} (\cite{jajora} and references therein)
has been singled out. In this short
letter we  prove that, actually, there is a very simple cylindrically symmetric 
dust universe included in the Szekeres family, which surprisingly has 
not been studied or considered hitherto. Some plausible reasons for this 
overlooking will also be analized: apparently, some minor errors in \cite{BOTO} 
led to a misunderstanding which has kept the solution ``behind the screen''.

As a matter of fact, the solution was recently rederived as the only dust
solution defined by the following separation Ansatz
in half-null coordinates \cite{Raultesi}:
\[
\d s^2=-2\d u \d v +\left[f(u)+g(v)\right]^2\d y^2 +
\left[h(u)+k(v)\right]^2\d z^2,
\]
where none of the derivatives of the given functions vanish,
and the coordinates are taken to be comoving,
that is, the dust velocity vector field reads
$\vec u\propto \partial/\partial u+\partial/\partial v$. Changing to 
typical Lorentzian coordinates the line-element takes the strikingly 
simple form
\be
\d s^2=-\d t^2+ \d \rho^2 + \rho^2 \d \pp^2+
\left(1-\frac{t^2+\rho^2}{\alpha^2} \right)^2\d z^2,
\label{eq:dustszek}
\ee
where $\alpha$ is an arbitrary constant. The cylindrical symmetry
is explicit, where $\rho > 0$ is the cylindrical radius (the axis is at
$\rho=0$) and $\pp$ is the corresponding angular coordinate running from
$0$ to $2\pi$. It can be checked that there are no further
isometries. The coordinate $z$ is taken to run from $-\infty$ to 
$\infty$, while the coordinate $t$ will be restricted by the singularity of
the spacetime, as we will see presently.

The dust is comoving in the coordinates of (\ref{eq:dustszek}),
its energy density given by
\[
\varrho=4\left(\alpha^2-t^2-\rho^2 \right)^{-1},
\]
while the non-vanishing components of the kinematical quantities for 
$\vec u=\partial/\partial t$
computed in the orthonormal co-basis given by
$\bm{\theta}^\alpha \propto \d x^\alpha$ are
\[
\theta=- 2t\left(\alpha^2-t^2-\rho^2\right)^{-1},
\]
\[
\sigma_{11}=\sigma_{22}=-\frac{\theta}{3},\hspace{1cm}
\sigma_{33}=2\frac{\theta}{3}.
\]
Of course, the acceleration and vorticity vanish.
The scalars of the Weyl tensor, computed in the null tetrad (see \cite{KRAM})
$\bm{k}=2^{-1/2}(\bm{\theta}^0-\bm{\theta}^1)$,
$\bm{l}=2^{-1/2}(\bm{\theta}^0+\bm{\theta}^1)$,
$\bm{m}=2^{-1/2}(\bm{\theta}^2+i \bm{\theta}^3)$,
read
\[
\Psi_0=\Psi_4=3\Psi_2=-\frac{1}{4}\varrho,\hspace{1cm}
\Psi_1=\Psi_3=0,
\]
and thus these solutions are of Petrov type D.

\begin{figure}[tb]
\centering
\begin{picture}(70,70)
\put(35,65){$t$}
\put(68
,30){$\rho$}
\mbox{\epsfig{file=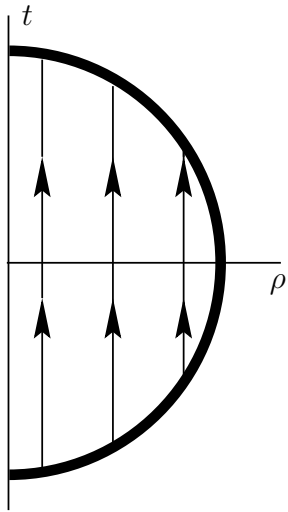,width=7cm}}
\end{picture}
\caption{This diagram corresponds to the $(t,\rho)$ plane of the spacetime
with line-element given by (\ref{eq:dustszek}). As usual,
null lines are at 45$^o$.
The whole spacetime is the product of this plane with the
$(\pp,z)$-surfaces, which correspond to the orbits of the
2-dimensional isometry group defining the cylindrical symmetry.
The thick curve represents the spacetime singularity. The dust 
particles move along the geodesics represented
by arrowed vertical lines in the physical region where $\varrho>0$,
whose ``size'' is measured by the radius $\alpha$ of the
semi-circle. In the limit with $\alpha \rightarrow \infty$ the 
singularity disappears and the solution becomes flat Minkowski 
spacetime.
}
\label{fig:szek}
\end{figure}

The expressions of the energy density and of the Weyl tensor show that a
curvature singularity appears at $t^2+\rho^2-\alpha^2\rightarrow 0$,
which can be viewed as a {\it connected} hypersurface.
The meaningful spacetime region (defined by $\varrho>0$) is clearly given by
\[
t^2+\rho^2<\alpha^2,
\]
and therefore the singularity `wraps' the entire manifold.
This singularity has a general character, as it is spacelike in the regions
given by
$t\in[-\alpha,-\alpha/\sqrt{2})\cup (\alpha/\sqrt{2},\alpha]$
(so that $\rho\in [0,\alpha/\sqrt{2})$), null in
$t=\pm \alpha/\sqrt{2}$ (where $\rho=\alpha/\sqrt{2}$),
and timelike at the region $t\in (-\alpha/\sqrt{2},\alpha/\sqrt{2})$
(thus $\rho\in (\alpha/\sqrt{2},\alpha]$).
This is shown in Figure \ref{fig:szek}.

All the causal geodesics
start and end at the singularity. As a particular case,
the timelike geodesics defined by the dust flow begin at the part of
the singularity where $t<0$ with $\theta\rightarrow \infty$, go on 
expanding up to $t=0$, where the dust starts to contract and 
eventually die in the future part of the
singularity ($t>0$) where $\theta \rightarrow -\infty$.
Actually, all endless causal curves passing through any given point
of the spacetime reach the singularity both in their past and future
for a finite value of their generalized affine parameter.
Thus, the singularity is past and future universal and can be termed as
a big-bang/big-crunch singularity with a general character
(see \cite{totxosing} for definitions).

Some interesting features can be deduced from the diagram of Figure 
\ref{fig:szek}. To start with, the big-bang and the big-crunch are
not simultaneous, something which happens in some other well-known cases,
see e.g.  \cite{Kra}. The total proper lifetime of any dust particle is 
always finite and depends 
on its position (on $\rho$), so that it is smaller for bigger $\rho$. 
This total proper time has no lower bound, so that one can find dust 
particles with as short a lifetime as desired. On the other hand, the maximun
lifetime for a dust particle is reached at the axis, and is given by 
$t_{max}=2\alpha$. This increases with $\alpha$. Notice that for very 
big $\alpha$ the metric approaches flat spacetime, and in the 
limit $\alpha \rightarrow \infty$ the spacetime is exactly Minkowskian 
(and the singularity disappears!).

Actually, for small values of $1/\alpha$ the metric can be 
considered as a small perturbation of the flat Minkowski metric. As 
$\alpha$ is a free parameter, the smallness of the perturbation can be 
chosen at will. Notice also that the spacetime is obviously globally 
hyperbolic and, furthermore, the hypersurface $t=0$ is maximal in the 
sense that its second fundamental form vanishes. Thus, this spacetime seem
to indicate that the classical results on the global non-linear stability
of Minkowski spacetime \cite{CK,KN} may not be generalizable to the case 
of {\it non-vacuum} perturbations (or to 
non-asymptotically-flat perturbations). It is interesting to see how 
this metric shows that the appearance of an arbitrarily small 
quantity of matter {\it everywhere} in flat spacetime destroys its 
global structure and encloses the matter within a strong curvature 
singularity.

Another curious property is that there are causal curves (and geodesics) 
which never cross the $t=0$ hypersurface, starting and ending at the 
part of the singularity with $t<0$. Thus, a part of the singularity 
with $t<0$ is actually the future singularity for some physical 
particles (not for the dust, though). In fact, some dust particles
(and other particles as well) can communicate with each other at their 
respective birth times. In other words, some dust particles can send 
causal signals (even from their bang starting time) which reach other 
dust particles at their corresponding birth times. Thus, for instance, 
a dust particle could tell another dust particle (with a big $\rho$)
about its fate, so that the latter may decide to leave the dust motion and 
travel to smaller values of $\rho$ in order to live longer. Similarly, 
a dust particle dying in the big-crunch can send a signal at that very 
moment showing its fate to other dust particles which could see how 
their end looks like. In other words, the big-crunch singularity 
is partly ``naked'' and visible for the dust. In general, and as the metric 
is time-symmetric, all statements for the future have their 
corresponding past counterpart.

The line-element (\ref{eq:dustszek}) is a dust solution with
Petrov type D, so that it belongs to
the family of the so-called Szekeres spacetimes
\cite{SZE,Kra,wwszec},
which were later generalized to a bigger family
of Petrov D spacetimes \cite{SZA,SZAWW,Kra} known
as the Szafron-Szekeres solutions.
The Szekeres solutions
do not contain any isometry in general \cite{BONSUTO}, but
particular families of this kind of inhomogeneous
solutions with a $G_2$ group of isometries were
given in \cite{BOTO} and \cite{WW1}.
More precisely, the metric (\ref{eq:dustszek}) is in fact
contained in the case PII of the parabolic models in the classification
of the Szekeres solutions presented by Bonnor and Tomimura in \cite{BOTO}
(page 86) where the evolution of these models was studied.

Nevertheless, this solution was not under the scope of the aforementioned
references mainly due to the fact that interest was focused
on the study of the solutions containing only an initial spacelike
singularity. Indeed, in \cite{BOTO} it was already established that the
Szekeres solutions contain two possible singularities, corresponding
to the vanishing of two functions appearing in the metric:
the so-called Friedmann function $R(t)$
and another arbitrary function $Q(t,x,y,z)$ (in what follows
we are using the notation of \cite{BOTO}).
The singularity coming from $R=0$ occurs on a spacelike hypersurface $t=0$,
whereas that from $Q=0$ admit a great variety of behaviours,
and thus in \cite{BOTO} the assumption that $t=0$ were reached
before $Q=0$ was tacitly demanded. The arbitrary functions ought then 
to be chosen such that
$Q>0$ in the spacetime in order to have $\varrho>0$.
However, in the PII class one has $R=1$ so that the only singularity
corresponds to $Q=0$.
As we have seen, the line-element (\ref{eq:dustszek}) explicitly reaches
$Q=0$, which now is simply $Q=1-(t^2+\rho^2)/\alpha^2$.
The rest of the functions
as they appear in \cite{BOTO} are given as follows: $\mu=\nu=\sigma=0$,
$\omega=1$, $\beta=\mbox{const.}(=-1/\alpha^2)$.
In fact, it was argued in \cite{BOTO} that the PII class would always 
contain a region of the spacetime with a negative energy density,
but this is not the case as we have explicitly proved for the solution
under consideration: for the solution (\ref{eq:dustszek}) the condition
$Q>0$ is enough to define a spacetime with $\varrho>0$ everywhere.

Regarding \cite{GOWA}, the class PII was again dismissed
because it has no non-vacuum
Friedmann-Lema\^{\i}tre-Robertson-Walker (FLRW) specialization, so that
its singularity structure, as well as the study of its evolution,
cannot be performed with a reformulated line-element in terms of
increasing and decreasing density perturbation modes of FLRW
dust models, which was the technique used.


\section*{Acknowledgements}
We are grateful to W.B. Bonnor, A. Krasi\'{n}ski, P. Szekeres and J. Wainwright
for their comments.
The authors also thank financial support from the Basque Country 
University under project number UPV 172.310-G02/99. RV thanks the
Spanish Secretar\'{\i}a de Estado
de Universidades, Investigaci\'on y Desarrollo,
Ministerio de Educaci\'on y Cultura, grant No. EX99 52155527.

\end{document}